%% file: main.tex
\documentclass[runningheads]{llncs}
\input{commands}
\usepackage{graphicx}
\begin{document}
\captionsetup[table]{skip=8pt}

\captionsetup[figure]{name={Fig. }, labelfont=bf, labelsep=period} 
\captionsetup[table]{name={Table}, labelfont=bf, labelsep=period}

\setlength{\floatsep}{4pt plus 4pt minus 1pt}
\setlength{\textfloatsep}{4pt plus 2pt minus 2pt}
\setlength{\intextsep}{4pt plus 2pt minus 2pt}
\setlength{\dbltextfloatsep}{3pt plus 2pt minus 1pt}
\setlength{\dblfloatsep}{3pt plus 2pt minus 1pt} 

\setlength{\abovecaptionskip}{3pt}
\setlength{\belowcaptionskip}{3pt}
\setlength{\abovedisplayskip}{3pt}
\setlength{\belowdisplayskip}{3pt}

\title{How Can Graph Neural Networks Help Document Retrieval: A Case Study on CORD19 with Concept Map Generation}
\titlerunning{Graph Neural Networks for Document Retrieval}
\author{Hejie Cui\orcidID{0000-0001-6388-2619} \and Jiaying Lu\orcidID{0000-0001-9052-6951} \and Yao Ge\orcidID{0000-0002-3323-7130} \and Carl Yang\orcidID{0000-0001-9145-4531}\thanks{Corresponding Author.}}
\authorrunning{H. Cui et al.}
%
\institute{Department of Computer Science, Emory University \\
\email{\{hejie.cui, jiaying.lu, yao.ge, j.carlyang\}@emory.edu}}
%
%
\maketitle              
\begin{abstract}
Graph neural networks (GNNs), as a group of powerful tools for representation learning on irregular data, have manifested superiority in various downstream tasks. With unstructured texts represented as concept maps, GNNs can be exploited for tasks like document retrieval. Intrigued by how can GNNs help document retrieval, we conduct an empirical study on a large-scale multi-discipline dataset CORD-19. Results show that instead of the complex structure-oriented GNNs such as GINs and GATs, our proposed semantics-oriented graph functions achieve better and more stable performance based on the BM25 retrieved candidates. 
Our insights in this case study can serve as a guideline for future work to develop effective GNNs with appropriate semantics-oriented inductive biases for textual reasoning tasks like document retrieval and classification. All code for this case study is available at \url{https://github.com/HennyJie/GNN-DocRetrieval}.

\end{abstract}
\input{sections/intro.tex}
\input{sections/methods.tex}
\input{sections/experiment.tex}
\input{sections/conclusion.tex}

%
%
%
\bibliographystyle{splncs04}
\bibliography{reference}
%





\end{document}

%% file: commands.tex
\usepackage{booktabs} 
\usepackage{amsfonts}
\usepackage{bm}
\usepackage{subcaption}
\usepackage{cite}
\usepackage{colortbl}
\usepackage{xcolor}
\newcommand{\header}[1]{{\vspace{+1mm}\flushleft \textbf{#1}}}
\usepackage{amsmath}
\usepackage{multirow}
\usepackage{caption}
\usepackage{url}

%% file: sections/intro.tex
\section{Introduction}
\label{sec:intro}
Concept map, which models texts as a graph with words/phrases as vertices and relations between them as edges, has been studied to improve information retrieval tasks previously \cite{DBLP:conf/cscwd/ZhangWXP18,DBLP:journals/prl/FarhiB18,DBLP:conf/ecir/Kamphuis20}. Recently, graph neural networks (GNNs) attract tremendous attention due to their superior power established both in theory and through experiments \cite{kipf2016semi,hamilton2017inductive,velivckovic2018graph,liu2019geniepath, DBLP:journals/corr/abs-2107-01495}. Empowered by the structured document representation of concept maps, it is intriguing to apply powerful GNNs for tasks like document classification \cite{DBLP:conf/sigir/YangZWL020} and retrieval \cite{DBLP:conf/aaai/ZhangZCWW21}. Take Fig. \ref{fig:toy-example} as an example. Towards the query  about ``violent crimes in society'', a proper GNN might be able to highlight query-relevant concept of ``crime'' and its connection to ``robbery'' and ``citizen'', thus ranking the document as highly relevant. On the other hand, for another document about precaution, the GNN can capture concepts like ``n95 mask'' and ``vaccine'', together with their connections to ``prevention'', thus ranking it as not so relevant.

\header{Present work.} In this work, we explore how GNNs can help document retrieval with generated concept maps. The core contributions are three-fold:
\begin{list}{$\bullet$}{\leftmargin=1em \itemindent=0em}
\item We use constituency parsing to construct semantically rich concept maps from documents and design quality evaluation for them towards document retrieval.
\item We investigate two types of graph models for document retrieval: the structure-oriented complex GNNs and our proposed semantics-oriented graph functions.
\item By comparing the retrieval results from different graph models, we provide insights towards GNN model design for textual retrieval, with the hope to prompt more discussions on the emerging areas such as IR with GNNs. 
\end{list}

%% file: sections/methods.tex
\section{GNNs for Document Retrieval}
\label{sec:methods}
\begin{figure*}[t]
    \centering
    \includegraphics[width=\linewidth]{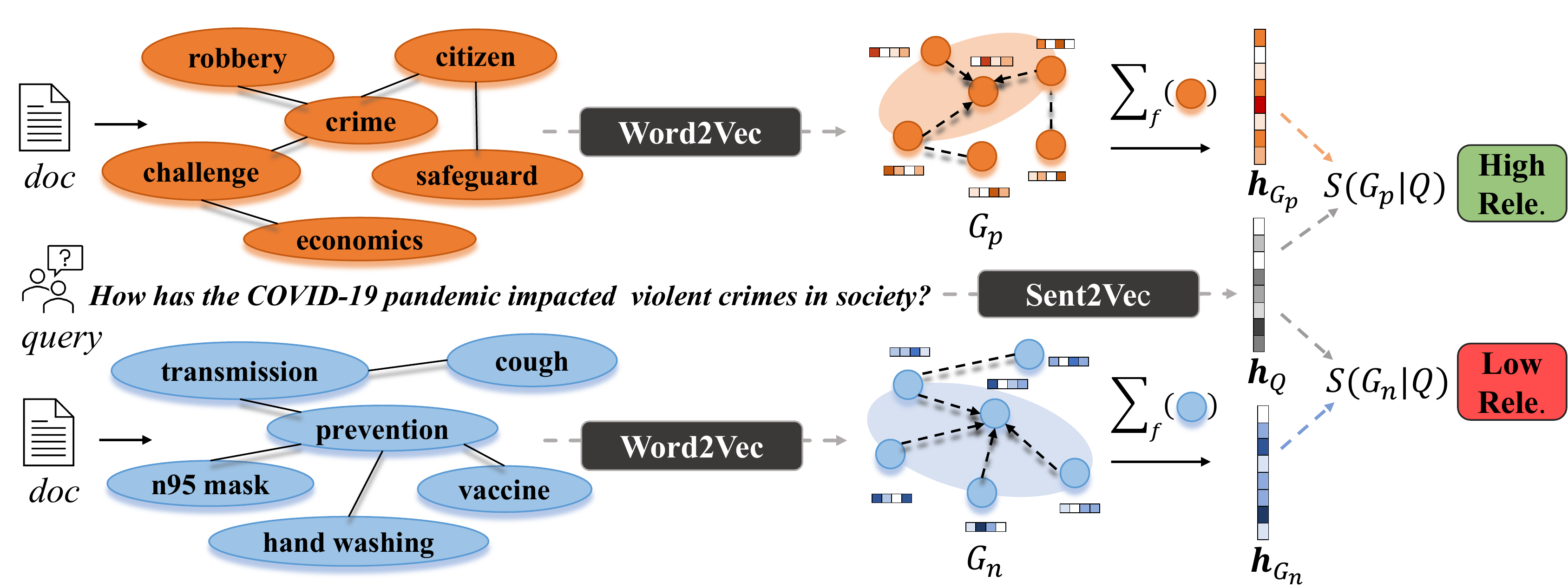}
    \caption{An overview of GNN-based document retrieval. 
    }
    \label{fig:toy-example}
\end{figure*}
\subsection{Overview}
In this section, we describe the process of GNN-based document retrieval. As is shown in Fig. \ref{fig:toy-example}, concept maps \(G = \{V, E\}\) are first constructed for documents. Each node \(v_i \in V\) is a concept (usually a word or phrase) in the document, associated with a frequency $f_i$ and an initial feature vector $\bm a_i$ from the pretrained model. The edges in \(E\) denote the interactions between concepts. GNNs are then applied to each individual concept map, where node representation \(\bm{h}_i \in \mathbb{R}^{d}\) is updated through neighborhood transformation and aggregation. The graph-level embedding \(\bm{h}_G \in \mathbb{R}^{d}\) is summarized over all nodes with a read-out function. 

For the training of GNN models, the widely-used triplet loss in retrieval tasks \cite {DBLP:conf/iccv/ManmathaWSK17,DBLP:conf/kdd/YingHCEHL18,DBLP:conf/kdd/YangPZP0RL20} is adopted. Given a triplet $(Q, G_p, G_n)$ composed by a relevant document $G_p$ (denoted as positive) and an irrelevant document $G_n$ (denoted as negative) to the query $Q$, the loss function is defined as:
\begin{equation}
\setlength{\abovedisplayskip}{3pt}
\setlength{\belowdisplayskip}{3pt}
L(Q, G_p, G_n)=\max \left\{ S(G_n \mid Q) - S(G_p \mid Q)+\textit{margin}, 0\right\}.
\label{eq:triplet}
\end{equation}
The relevance score $S \left(G \mid Q \right)$ is calculated as $ \frac{\bm{h}_G \cdot \bm{h}_Q}{\|\bm{h}_G \|\|\bm{h}_Q\|}$, where $\bm h_G$ is the learned graph representation from GNN models and $\bm h_Q$ is the query representation from a pretrained model.
In the training process, the embeddings of relevant documents are pulled towards the query representation, whereas those of the irrelevant ones are pushed away. For retrieval in the testing phrase, documents are ranked according to the learned relevance score $S(G \mid Q)$.

\subsection{Concept Maps and Their Generation}
\label{ssec:concept-map}
Concept map generation, which aims to distill structured information hidden under unstructured text and represent it with a graph, has been studied extensively in literature \cite{DBLP:journals/ce/ChenKWC08,DBLP:conf/nips/YangZSL019,DBLP:conf/wsdm/YangZWLKWX020,DBLP:conf/aaai/ZhangZCWW21}. 
Since entities and events often convey rich semantics, they are widely used to represent core information of documents \cite{christensen-etal-2013-towards,li-etal-2020-connecting,lu2021evaluation}. 
However, according to our pilot trials, existing concept map construction methods based on name entity recognition (NER) or relation extraction (RE) often suffer from limited nodes and sparse edges.
Moreover, these techniques rely on significant amounts of training data and predefined entities and relation types, which restricts the semantic richness of the generated concept maps \cite{wang2018comparative}.

To increase node/edge coverage, we propose to identify entities and events by POS-tagging and constituency parsing \cite{manning-etal-2014-stanford}. Compared to concept maps derived from NER or RE, our graphs can identify more sufficient phrases as nodes and connect them with denser edges, since pos-tagging and parsing are robust to domain shift \cite{mcclosky2010automatic,yu2015domain}.
The identified phrases are filtered via articles removing and lemmas replacing, and then merged by the same mentions. To capture the interactions (edges in graphs) among extracted nodes, we follow the common practice in phrase graph construction \cite{mihalcea2004textrank,rose2010automatic,krallinger2005sentence} that uses the sliding window technique to capture node co-occurrence. The window size is selected through grid search. Our proposed constituency parsing approach for concept map generation alleviates the limited vocabulary problem of existing NER-based methods, thus bolstering the semantic richness of the concept maps for retrieval. 

\subsection{GNN-based Concept Map Representation Learning}
\label{ssec:gnn}

\header{Structure-oriented complex GNNs}
\label{ssec:sota}
Various GNNs have been proposed for graph representation learning \cite{kipf2016semi,hamilton2017inductive,xu2018powerful,velivckovic2018graph}. The discriminative power of complex GNNs mainly stems from the 1-WL test for graph isomorphism, which exhaustively capture possible graph structures so as to differentiate non-isomorphic graphs \cite{xu2018powerful}.
To investigate the effectiveness of structured-oriented GNNs towards document retrieval, we adopt two state-of-the-art ones, Graph isomorphism network (GIN) \cite{xu2018powerful} and Graph attention network (GAT) \cite{velivckovic2018graph}, as representatives. 



\header{Semantics-oriented permutation-invariant graph functions}
\label{ssec:simple}
The advantage of complex GNNs in modelling interactions may become insignificant for semantically important task.
In contrast, we propose the following series of graph functions oriented from semantics perspectives.
\begin{itemize}
    \item \textbf{N-Pool}: independently process each single node $v_i$ in the concept map by multi-layer perceptions and then apply a read-out function to aggregate all node embeddings $\bm a_{i}$ into the graph embedding $\bm h_{G}$, i.e., 
    \begin{small}
    \begin{equation}
    \setlength{\abovedisplayskip}{3pt}
    \setlength{\belowdisplayskip}{3pt}
    \bm h_{G}=\operatorname{READOUT}\Big( \{\operatorname{MLP}(\bm a_{i}) \mid v_i \in V \}\Big).
    \end{equation}
    \end{small}
    \item \textbf{E-Pool}: for each edge $e_{ij}=(v_i, v_j)$ in the concept map, the edge embedding is obtained by concatenating the projected node embedding $\bm a_{i}$ and $\bm a_{j}$ on its two ends to encode first-order interactions, i.e., 
    \begin{small}
    \begin{equation}
    \setlength{\abovedisplayskip}{3pt}
    \setlength{\belowdisplayskip}{3pt}
    \bm h_{G}=\operatorname{READOUT}\Big(\left\{cat \left( \operatorname{MLP}(\bm a_{i}), \operatorname{MLP}(\bm a_{j}) \right) \mid e_{ij} \in E\right\}\Big).
    \end{equation}
    \end{small}
    \item \textbf{RW-Pool}: for each sampled random walk $p_i = (v_1, v_2, \ldots, v_m)$ that encode higher-order interactions among concepts ($m=2,3,4$ in our experiments), the embedding is computed by the sum of all node embeddings on it, i.e.,
    \begin{small} 
    \begin{equation}
    \setlength{\abovedisplayskip}{3pt}
    \setlength{\belowdisplayskip}{3pt}
    \bm h_{G}=\operatorname{READOUT}\Big( 
    \{ sum \left( \operatorname{MLP}(\bm a_1), \operatorname{MLP}(\bm a_2), \ldots, \operatorname{MLP}(\bm a_m) \right ) \mid p_i \in P \}\Big).
    \end{equation}
    \end{small}
\end{itemize}
All of the three proposed graph functions are easier to train and generalize. They preserve the \textit{message passing} mechanism of complex GNNs \cite{gilmer2017neural}, which is essentially \textit{permutation invariant} \cite{maron2019universality,maron2018invariant,keriven2019universal}, meaning that the results of GNNs are not influenced by the orders of nodes or edges in the graph; while focusing on the basic semantic units and different level of interactions between them.

%% file: sections/experiment.tex
\section{Experiments}
\begin{table}[t]
\centering
\caption{The similarity of different concept map pairs.}
\resizebox{0.7\textwidth}{!}{
\begin{tabular}{cccccc}
\toprule
Pair Type & \# Pairs & \bf NCR ($\%$)& \bf NCR+ ($\%$)& \bf ECR ($\%$) & \bf ECR+ ($\%$)\\ \midrule
Pos-Pos & 762,084  & 4.96 & 19.19 & 0.60 &  0.78       \\
Pos-Neg & 1,518,617 & 4.12 & 11.75 & 0.39 &  0.52       \\
\rowcolor{gray!20} (\textit{t-score}) & -  & (\textit{187.041}) & (\textit{487.078}) & (\textit{83.569}) &  (\textit{105.034})    \\
Pos-BM & 140,640 & 3.80 & 14.98 & 0.37 & 0.43 \\
\rowcolor{gray!20} (\textit{t-score}) & - & (\textit{126.977}) & (\textit{108.808}) & (\textit{35.870})  & (\textit{56.981}) \\
\bottomrule
\end{tabular}
}
\label{tab:structure}
\end{table}

\label{sec:exp}

\subsection{Experimental Setup}
\header{Dataset} We adopt a large scale multi-discipline dataset  from the TREC-COVID\footnote{\label{ft:trec-covid}https://ir.nist.gov/covidSubmit/} challenge \cite{DBLP:journals/jbi/RobertsABDLSVWH21} based on the CORD-19\footnote{\label{ft:cord19}https://github.com/allenai/cord19} collection \cite{wang-etal-2020-cord}. 
The raw data includes a corpus of 192,509 documents from broad research areas, 50 queries about the pandemic that interest people, and 46,167 query-document relevance labels. 
\header{Experimental settings and metrics} 
We follow the common two-step practice for the large-scale document retrieval task \cite{liu2011learning,dang2013two,nogueira2019passage}. The initial retrieval is performed on the whole corpus with full texts through BM25 \cite {Robertson1994OkapiAT}, a traditional yet widely-used baseline. In the second stage, we further conduct re-ranking on the top 100 candidates using different graph models. 
The node features and query embeddings are initialized with pretrained models from \cite{zhang2019biowordvec, chen2019biosentvec}.
NDCG@20 is adopted as the main evaluation metric for retrieval, which is used for the competition leader board. Besides NDCG@\textit K, we also provide Precision@\textit K and Recall@\textit K (\textit K=10, 20 for all metrics).

\subsection{Evaluation of Concept Maps}

We empirically evaluate the quality of concept maps generated from Section \ref{ssec:concept-map}. The purpose is to validate that information in concept maps can indicate query-document relevance, and provide additional discriminative signals based on the initial candidates.
Three types of pairs are constructed: a Pos-Pos pair consists of two documents both relevant to a query; a Pos-Neg pair consists of a relevant and an irrelevant one; and a Pos-BM pair consists of a relevant one and a top-20 one from BM25. 
Given a graph pair $G_i$ and $G_j$, their similarity is calculated via four measures:
the node coincidence rate (NCR) defined as $\frac{\lvert V_i \cap V_j \rvert}{\lvert V_i \cup V_j \rvert}$;
NCR+ defined as NCR weighted by the tf-idf score \cite{baeza1999modern} of each node;
the edge coincidence rate (ECR) where an edge is coincident when its two ends are contained in both graphs;
and ECR+ defined as ECR weighted by the tf-idf scores of both ends.

It is shown in Table \ref{tab:structure} that Pos-Neg pairs are less similar than Pos-Pos under all measures, indicating that concept maps can effectively reflect document semantics. Moreover, Pos-BM pairs are not close to Pos-Pos and even further away than Pos-Neg. This is because the labeled ``irrelevant'' documents are actually hard negative ones difficult to distinguish. Such results indicate the potential for improving sketchy candidates with concept maps. 
Besides, student's t-Test\cite{hogg2005introduction} is performed, where standard critical values of (Pos-Pos, Pos-Neg) and (Pos-Pos, Pos-BM) under 95\% confidence are 1.6440 and 1.6450, respectively. The calculated \textit{t-scores} shown in Table \ref{tab:structure} strongly support the significance of differences.

\begin{table}[t]
\centering
\caption{\label{tab:perform}The retrieval performance results of different models.}
\resizebox{0.7\textwidth}{!}{
\begin{tabular}{cccccccc}
\toprule
\multirow{2.5}{*}{Type} & \multirow{2.5}{*}{Methods} & 
\multicolumn{2}{c}{\bf Precision ($\%$)}&\multicolumn{2}{c}{\bf Recall ($\%$)}&\multicolumn{2}{c}{\bf NDCG ($\%$)}\\
\cmidrule(lr){3-4} \cmidrule(lr){5-6} \cmidrule(lr){7-8} 
& & {\it k=$10$} & {\it k=$20$} & {\it k=$10$} & {\it k=$20$} & {\it k=$10$} & {\it k=$20$} \\
\midrule
\multirow{2}{*}{Traditional}
& BM25 & 55.20 & 49.00 & 1.36 & 2.39 & 51.37 & 45.91 \\
& Anserini & 54.00 & 49.60 & 1.22 & 2.25 & 47.09 & 43.82 \\
\midrule
\multirow{2}{*}{Structure-Oriented}
& GIN & 35.24 & 34.36 & 0.77 & 1.50 & 30.59 & 29.91 \\
& GAT & 46.48 & 43.26 & 1.08 & 2.00 & 42.24 & 39.49 \\
\midrule 
\multirow{3}{*}{Semantics-Oriented}
& \cellcolor{lightgray!20} N-Pool & \cellcolor{lightgray!20} 58.24 & \cellcolor{lightgray!20} 52.20 & \cellcolor{lightgray!20} 1.38 & \cellcolor{lightgray!20} 2.41 & \cellcolor{lightgray!20} 53.38 & \cellcolor{lightgray!20} 48.80 \\
& \cellcolor{lightgray!20} E-Pool & \cellcolor{lightgray!20} 59.60 & \cellcolor{lightgray!20} 53.88  & \cellcolor{lightgray!20} 1.40 & \cellcolor{lightgray!20} 2.49 & \cellcolor{lightgray!20} 56.11 & \cellcolor{lightgray!20} 51.16 \\
& \cellcolor{lightgray!20} RW-Pool &  \cellcolor{lightgray!20} \bf 59.84 & \cellcolor{lightgray!20} \bf 53.92 &  \cellcolor{lightgray!20} \bf 1.42 & \cellcolor{lightgray!20} \bf 2.53 & \cellcolor{lightgray!20} \bf 56.19 & \cellcolor{lightgray!20} \bf 51.41 \\
\bottomrule
\end{tabular}
}
\end{table}
\subsection{Retrieval Performance Results}
In this study, we focus on the performance improvement of GNN models based on sketchy candidates. Therefore, two widely-used and simple models, the forementioned BM25 and Anserini\footnote{https://git.uwaterloo.ca/jimmylin/covidex-trec-covid-runs/-/tree/master/round5, which is recognized by the competition organizers as a baseline result.}, are adopted as baselines, instead of the heavier language models such as BERT-based \cite{DBLP:conf/naacl/DevlinCLT19,DBLP:conf/emnlp/YilmazWYZL19,DBLP:journals/corr/abs-2007-12603} and learning to rank (LTR)-based \cite{DBLP:conf/icml/BurgesSRLDHH05,DBLP:journals/ir/WuBSG10} ones. The retrieval performance are shown in Table \ref{tab:perform}. All the values are reported as the averaged results of five runs under the best settings.

For the structure-oriented GIN and GAT, different read-out functions including mean, sum, max and a novel proposed tf-idf (i.e., weight the nodes using the tf-idf scores) are experimented, and tf-idf achieves the best performance. It is shown that GIN constantly fails to distinguish relevant documents while GAT is relatively better. However, they both fail to improve the baselines. This performance deviation may arise from the major inductive bias on complex structures, which makes limited contribution to document retrieval and is easily misled by noises. 
In contrast, our three proposed semantics-oriented graph functions yield significant and consistent improvements over both baselines and structure-oriented GNNs. Notably, E-Pool and RW-Pool improve the document retrieval from the initial candidates of BM25 by 11.4\% and 12.0\% on NDCG@20, respectively. Such results demonstrate the potential of designing semantics-oriented GNNs for textual reasoning tasks such as classification, retrieval, etc. 

\subsection{Stability and Efficiency}
We further examine the stability and efficiency of different models across runs.
\begin{figure}[t]
	\centering
	\subfloat[Stability comparison]{
	    \centering
		\includegraphics[width=0.35\linewidth]{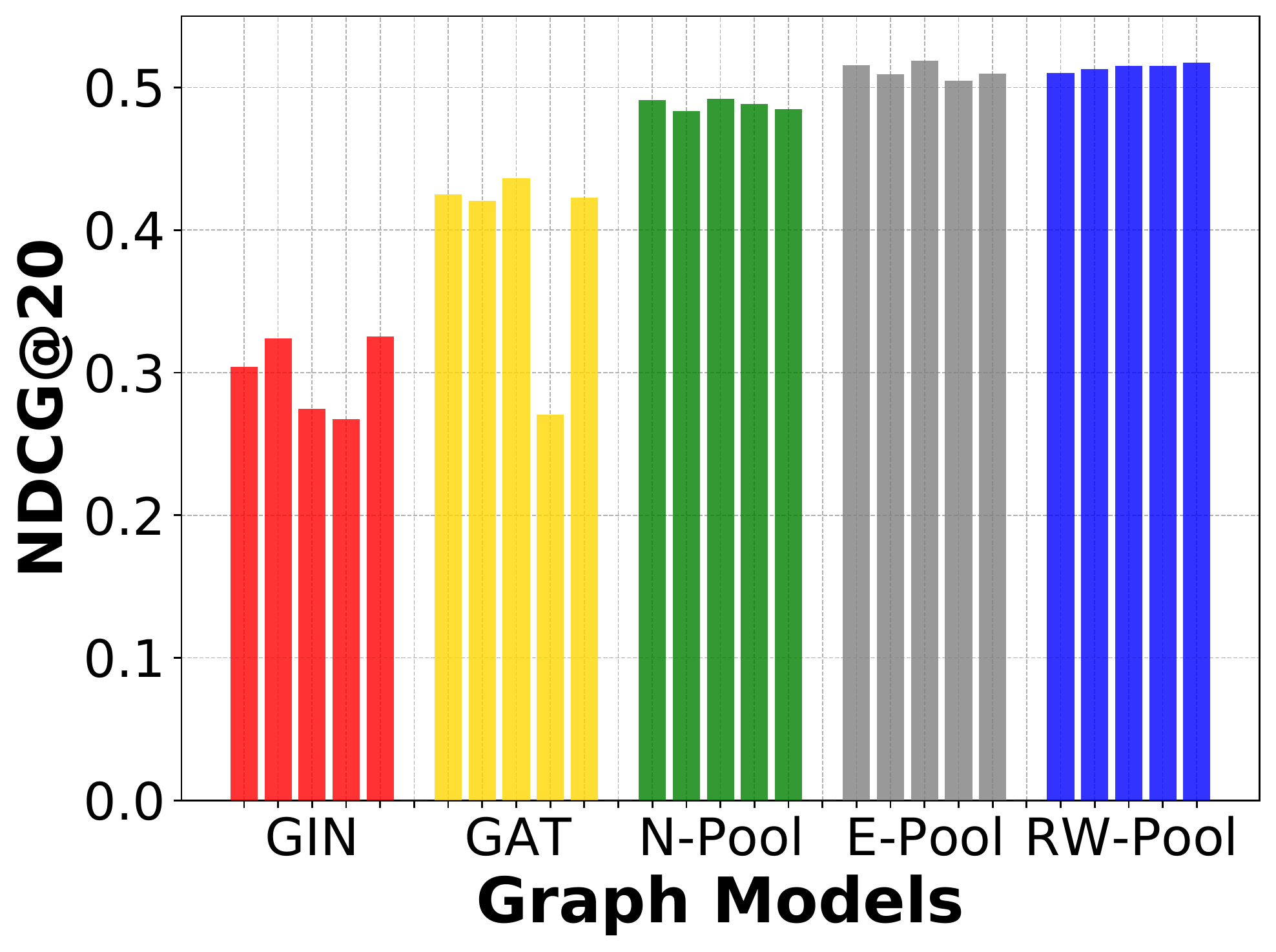}
		\label{fig:consis_ndcg20}
	}
	\subfloat[Efficiency comparison]{
	    \centering
		\includegraphics[width=0.35\linewidth]{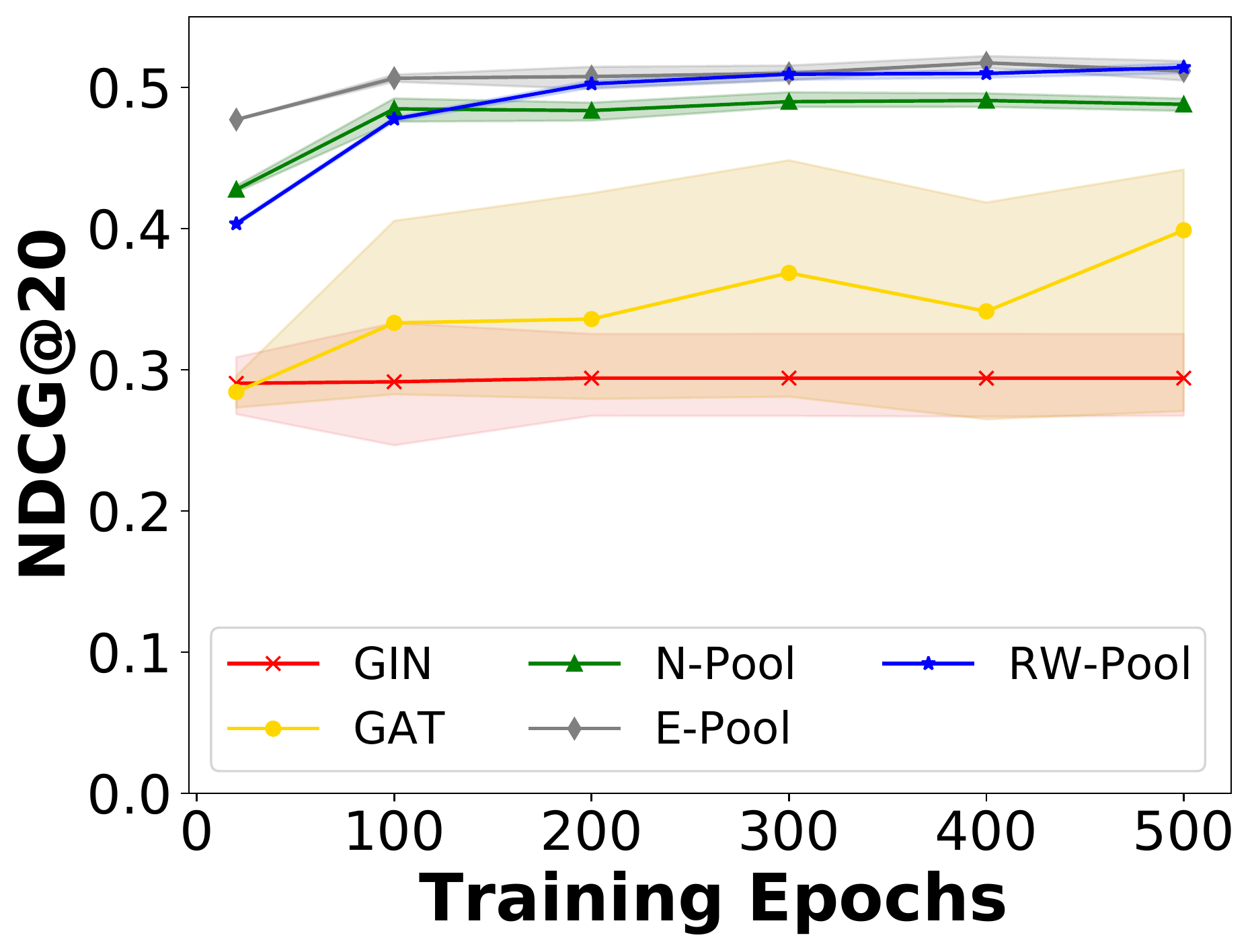}
		\label{fig:train_curve_ndcg20}
	}
	\caption{Stability and efficiency comparison of different graph models.}
	\label{fig:analysis}
\end{figure}
As is shown in Fig. \ref{fig:analysis}(a), GIN and GAT are less consistent, indicating the difficulty in training over-complex models. The training efficiency in Fig. \ref{fig:analysis}(b) shows that GIN can hardly improve during training, 
while GAT fluctuates a lot and suffers from overfitting.
In contrast, our proposed semantics-oriented functions perform more stable in Fig. \ref{fig:analysis}(a), and improve efficiently during training in Fig. \ref{fig:analysis}(b), demonstrating their abilities to model the concepts and interactions important for the retrieval task. Among the three graph functions, E-Pool and RW-Pool are consistently better than N-Pool, revealing the utility of simple graph structures. 
Moreover, RW-Pool converges slower but achieves better and more stable results in the end, indicating the potential advantage of higher-order interactions. 

%% file: sections/conclusion.tex
\section{Conclusion}
\label{sec:conclu}
In this paper, we investigate how can GNNs help document retrieval through a case study. Concept maps with rich semantics are generated from unstructured texts with constituency parsing.
Two types of GNNs, structure-oriented complex models and our proposed semantics-oriented graph functions are experimented and the latter achieves consistently better and stable results, demonstrating the importance of semantic units as well as their simple interactions in GNN design for textual reasoning tasks like retrieval.
In the future, more textual datasets such as news, journalism and downstream tasks can be included for validation. Other types of semantics-oriented graph functions can also be designed based on our permutation-invariant schema, such as graphlet based-pooling. 